\documentclass[3p,times,procedia]{elsarticle}
\usepackage{color}

\usepackage{amssymb}
\usepackage{amsmath}

\makeatletter
\newcommand{\xRightarrow}[2][]{\ext@arrow 0359\Rightarrowfill@{#1}{#2}}
\makeatother

\usepackage[figuresright]{rotating}

\begin{document}

\begin{frontmatter}

\title{Nearly isentropic flow at sizeable $\eta/s$ }

\author{Aleksi Kurkela$^{1,2}$, Urs Achim Wiedemann$^{1}$ and Bin Wu$^{1}$}

\address{$^1$ Theoretical Physics Department, CERN, CH-1211 Gen\`eve 23, Switzerland\\
$^2$ Faculty of Science and Technology, University of Stavanger, 4036 Stavanger, Norway}

\begin{abstract}
Non-linearities in the harmonic spectra of hadron-nucleus and nucleus-nucleus collisions provide evidence for the dynamical response to 
azimuthal spatial eccentricities. Here, we demonstrate within the framework of transport theory that even the mildest 
interaction correction to a picture of free-streaming particle distributions, namely the inclusion of one perturbatively 
weak interaction (``one-hit dynamics''), will generically give rise to all observed linear and non-linear structures. We further argue 
that transport theory naturally accounts within the range of its validity for realistic signal sizes of the linear and
non-linear response coefficients observed in azimuthal momentum anisotropies with a large mean free path of the order
of the system size in peripheral ($\sim 50 \%$ centrality) PbPb or central pPb collisions. As a non-vanishing mean free path
is indicative of non-minimal dissipation, this challenges the perfect fluid paradigm of ultra-relativistic
nucleus-nucleus and hadron-nucleus collisions. 
\end{abstract}


\end{frontmatter}
\vspace{0.5cm}

{\bf Introduction}.
The applicability of relativistic transport theory ranges from the description of free-streaming particle distributions via weakly interacting systems of long mean free path up to systems exhibiting viscous fluid dynamic behavior. In this way, transport theory encompasses 
within a unified framework a broad range of candidate theories for the collective flow-like behavior observed in ultra-relativistic nucleus-nucleus~\cite{ALICE:2011ab} and, more recently, proton-nucleus~\cite{Abelev:2014mda,Khachatryan:2015waa,Aaboud:2017acw}, 
deuterium-nucleus~\cite{Adare:2014keg,Adamczyk:2015xjc} and high multiplicity proton-proton collisions~\cite{Khachatryan:2016txc,Aad:2015gqa}. 
 
 Arguably the best-studied candidate theory for collective dynamics in nuclear collisions is the viscous fluid dynamic limit of kinetic 
 transport. Fluid dynamic simulations of
 ultra-relativitic heavy ion collisions at RHIC and at the LHC 
 can account for hadronic spectra in nucleus-nucleus collisions, their azimuthal anisotropies $v_m$, their $p_T$-differential hadrochemical 
 composition, as well as for correlations between different $v_m$'s, see Refs.~\cite{Heinz:2013th,Romatschke:2017ejr} for recent reviews.
 Generically, these fluid dynamic simulations favor an almost perfect fluid with a very small ratio of shear viscosity to entropy density $\eta/s \sim {\cal O}(1/4\pi)$, suggesting 
 that the system is so strongly interacting that the notion of quasi-particles and mean free path becomes doubtful since no particle excitation propagates over distances 
 larger than ${\cal O}(1/T)$. However, up to finer differences~\cite{Adam:2015bka,Bzdak:2014dia} that may or may not be improvable via  
 tuning of model parameters,
 the main signal sizes and kinematic dependencies of $v_m$'s in nucleus-nucleus collisions have also 
 been accounted for~\cite{Xu:2011jm} by A-Multiphase-Transport-Model (AMPT)~\cite{Lin:2004en},  
 a simulation code of partonic and hadronic transport whose applicability relies on a sufficiently large mean free path. The supposed dichotomy between a strong coupling picture (based on negligible 
 mean free path and almost perfect fluidity) and a weak coupling picture based on spatially well-separated interactions is further challenged by the apparent
 phenomenologically compatibility of both descriptions with smaller collision systems.  In particular, exploratory 
 fluid dynamic simulations of pPb collisions have predicted ~\cite{Bozek:2011if,Bozek:2013uha}
 for an almost perfect fluid ($\eta/s \sim (1 - 2)/4\pi)$) the correct signal size and $p_T$-dependence of 
 anisotropic elliptic and triangular flow in pPb prior to data taking at the LHC~\cite{Abelev:2014mda,Khachatryan:2015waa,Aaboud:2017acw}, while the AMPT code provides a phenomenologically reasonable description of pPb data at LHC~\cite{Bzdak:2014dia} and
 smaller collision systems at RHIC~\cite{Koop:2015wea} with an apparently dilute system from which 
 approximately $\sim 50\%$ of all partons escape without rescattering~\cite{He:2015hfa} 
 (numbers quoted for d+Au collisions at RHIC). The discovery of heavy ion like behavior in smaller (pp and pPb) collision systems at the LHC
 and its confirmation at RHIC energies thus raises the question of whether the matter produced in these collisions shares properties of an almost perfect fluid, or whether the observed signatures of collectivity can arise in a system
 of particle-like excitations of significant mean free path. 
 
 Increasing $\eta/s$ in fluid dynamic simulations i) reduces the efficiency of translating spatial gradients into momentum anisotropies and ii) increases entropy production.
 Within a model parameter space that includes uncertainties in the initial conditions, one may imagine to increase $\eta/s$ by compensating effect i) with an increased
 initial density (and thus with increased initial spatial pressure gradients within a fixed initial spatial overlap area). But entropy increases with increasing initial density or $\eta/s$, and 
 any such variation of model parameters is therefore tightly constrained by event multiplicity.  This illustrates that a phenomenologically valid collective dynamics 
 needs to combine a nearly isentropic dynamics with an efficient mechanism for translating spatial anisotropies into momentum anisotropies. Remarkably, in the opposite limiting cases
 of a (nearly) perfect liquid and of a (nearly) free-streaming gas of particles, kinetic transport theory gives rise to a (nearly) isentropic 
 dynamics. The first limit is clearly realized in fluid dynamic simulations that are known to translate efficiently spatial into momentum anisotropies. At face value, the fact that transport codes like AMPT and MPC~\cite{Molnar:2000jh}
can build up an efficient collective dynamics with very few spatially separated collisions indicates that they dynamically realize the opposite limit of a nearly isentropic evolution close to free-streaming. This has the potential for a shift away from the perfect fluid paradigm 
and it thus deserves to be understood in detail.
 
 For any sufficiently complex simulation, the code {\it is} the model in the sense that the code is more than a tool for solving an easily stated set of equations of motion. For instance, fluid dynamic simulations do not code only for a solution
 to the viscous fluid dynamic equations of motion, but they interface those with a model of initial conditions, with a model for hadronization and with a hadronic rescattering phase. This necessary phenomenological complexity of a simulation can obscure the relation between phenomenological success and the physical mechanism at work.  For state-of-the-art fluid dynamic codes, however, the fluid dynamic part of the simulation is benchmarked against analytically known results of viscous Israel-Stewart theory, and many
 kinematic dependencies and measurable properties of $v_m$'s and their correlations can be understood at least qualitatively by solving in isolation for the viscous fluid dynamic equations of 
 motion with suitable initial conditions. It is the qualitative and semi-quantitative consistency between full phenomenological simulations and benchmarked solutions to an unambiguously defined set of 
 equations of motion that gives strength to the phenomenological conclusion of fluid dynamic behavior in heavy ion collisions.  
 
 For transport theory close to the free streaming limit, a corresponding program of anchoring major dynamical stages of full-fledged simulations on unambiguously and analytically defined benchmark calculations is largely missing (exceptions include 
early tests of $2\to 2$ parton cascades between the free-streaming and hydrodynamic limits~\cite{Gyulassy:1997ib,Zhang:1998tj},
 more recent comparisons of kinetic theory to fluid dynamics~\cite{Gabbana:2017uvc,Bouras:2010hm,Kurkela:2015qoa}).
What is at stake is to understand from simple physical principles and beyond any specific model implementation whether a 
particular manifestation of collective behavior is a {\it generic} property of transport theory. 
  Motivated by the conceivable phenomenological relevance of transport close to the free-streaming limit for ultra-relativistic hadronic collision systems, the aim of this paper is to pursue such a programme of establishing benchmark results, and to determine which characteristic signatures of collectivity will arise generically in such a framework.  
 
 {\bf The model}. We start from massless kinetic transport of a distribution $f(\vec{x}_\perp,\vec{p},\tau)$,
  assuming longitudinal boost invariance and focussing
 on the slice of central spatial rapidity $\eta_s = 0$ \cite{Baym:1984np}, 
 \begin{equation}
\partial_\tau f + \vec{v}_\perp \cdot \partial_{\vec{x}_\perp }f - \frac{p^z}{\tau} \partial_{p^z}f = -C[f]\,.
\label{eq1}
\end{equation}
We denote space-time coordinates by $\vec{x} = \left(x, y, z\right)$, $\vec{x}_\perp = \left( x, y\right)$ and normalized momenta by $v_\mu \equiv p_\mu/p$ with $p_\mu\, p^\mu = 0$
and $v^0 = 1$. 
 We are mainly interested in $p$-integrated distribution functions
$F(\vec x_\perp,\Omega,\tau) = \int\frac{4 \pi p^2 d p}{(2\pi)^3} p f$ that satisfy
\begin{equation}
\partial_\tau F + \vec{v}_\perp \cdot \partial_{\vec{x}_\perp} F - \frac{1}{\tau}v_z(1-v_z^2) \partial_{v_z} F + \frac{4 v_z^2}{\tau}F = -C[F] \, ,\label{eq2}
\end{equation}
where $\Omega$ is the solid angle in the momentum space.
 In order to derive the above equation from eq. (\ref{eq1}), we have taken $f$ as a function of $p, \phi$ and $v_z$ and we have used $\vec{v}_\perp=\sqrt{1-v_z^2}(\cos\phi,\sin\phi)$.
The collision kernel depends on the microscopic details of the interactions of the constituents. However, it is a generic property of interactions that they bring the system towards isotropy, and hence we consider a model --- \emph{isotropization-time approximation} --- in which the scattering between particles in a given volume element is assumed to distribute the particles isotropically in the rest frame of that given 
volume element
\begin{equation}
-C[F] = -\gamma \varepsilon^{1/4}(x) [-v_\mu u^{\mu} ] (F - F_{\rm iso})\, ,
\label{eq3}
\end{equation}
where $F_{\rm iso}$ is the isotropic distribution in the rest frame $u^\mu$. We further assume that the system is conformally symmetric such that the time scale for the interactions is proportional to the energy density scale of the medium, $l_{\rm mfp} \sim \left(\gamma \varepsilon^{1/4}\right)^{-1}$, where $\gamma$ is our single model parameter setting the isotropization rate. While in more realistic models, different momentum scales isotropize on different timescales~\cite{Arnold:2002zm,Kurkela:2017xis}, the momentum integral in the definition of $F$ is dominated by a single scale, whose isotropization time corresponds to our model parameter (see e.g.~\cite{Heller:2016rtz} for a comparison of this model to QCD effective kinetic theory, finding $\sim 10\%$ differences in relaxation of $T^{\mu\nu}$ when $\eta/s$ of the two models are matched). This caveat in mind, we shall restrain ourselves from making $p_T$-differential statements which would depend on details of the collision kernel that are not captured by a single isotropization timescale. 

The local rest frame energy density $\varepsilon$ and the rest frame velocity $u_\mu$ with normalization $u^\mu u_\mu = -1$ are defined by the Landau matching condition,
$u^\mu T_{\mu}^{\, \, \,\nu}  = - \varepsilon u^\nu$, with the energy-momentum tensor defined by the second velocity moments of $F$. The form of $F_{\rm iso}$ can be found by demanding that the unintegrated isotropic distribution $f_{\rm iso}$  depends only on the Lorentz scalar $p^\mu u_\mu$, 
$f_{\rm iso} = g(p^\mu u_\mu) $. Even though the exact functional dependence $g$ is not known for a non-equilibrium system, the integrated isotropic distribution can be 
determined explicitly from the dimensionality of the integrand and the requirement $\int \frac{d^3 p}{(2\pi)^3} p^\mu u_\mu (f - f_{iso})=0$ of local energy conservation, 
\begin{equation}
F_{iso}(\vec x_\perp,\Omega,\tau) = \int \frac{ 4\pi d p \, p^3 }{(2\pi)^3}   g(p u^\mu v_\mu)  = \frac{\varepsilon(\vec x_\perp,\tau)}{(-u_\mu v^{\mu})^4}\, . 
\label{eq4}
\end{equation}
This isotropization time approximation is closely related to the usual relaxation time approximation, and
in the hydrodynamical limit,  its transport coefficients are $\tau_\pi = (\gamma \varepsilon^{1/4})^{-1}$ and kinetic shear viscosity 
$\frac{\eta}{\varepsilon+P} = \frac{\eta}{s T} =  \frac{1 }{\gamma \epsilon^{1/4}}\frac{1}{ 5}$. While kinetic theory (\ref{eq1}) 
does not depend on the relation $\epsilon = a T^4$  between energy density and temperature, this relation enters if one
wants to express the only model parameter $\gamma$ in terms of the shear viscosity over entropy ratio. For $a \approx 13$
consistent with QCD lattice results, we find 
\begin{equation}
\frac{\eta}{s} \approx \frac{0.11}{\gamma}\, . 
\label{eq5}
\end{equation}

\begin{figure}
 \includegraphics[width=0.22\textwidth]{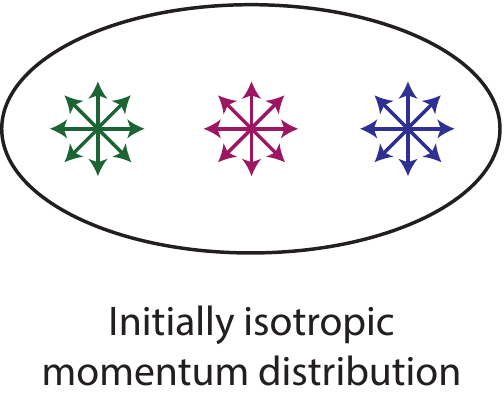} 
 \hfill
 \includegraphics[width=0.35\textwidth]{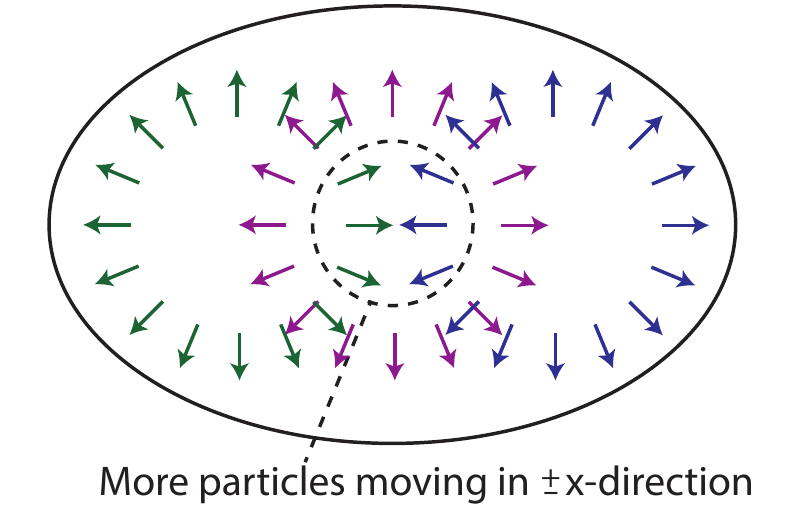}
  \hfill
  \includegraphics[width=0.35\textwidth]{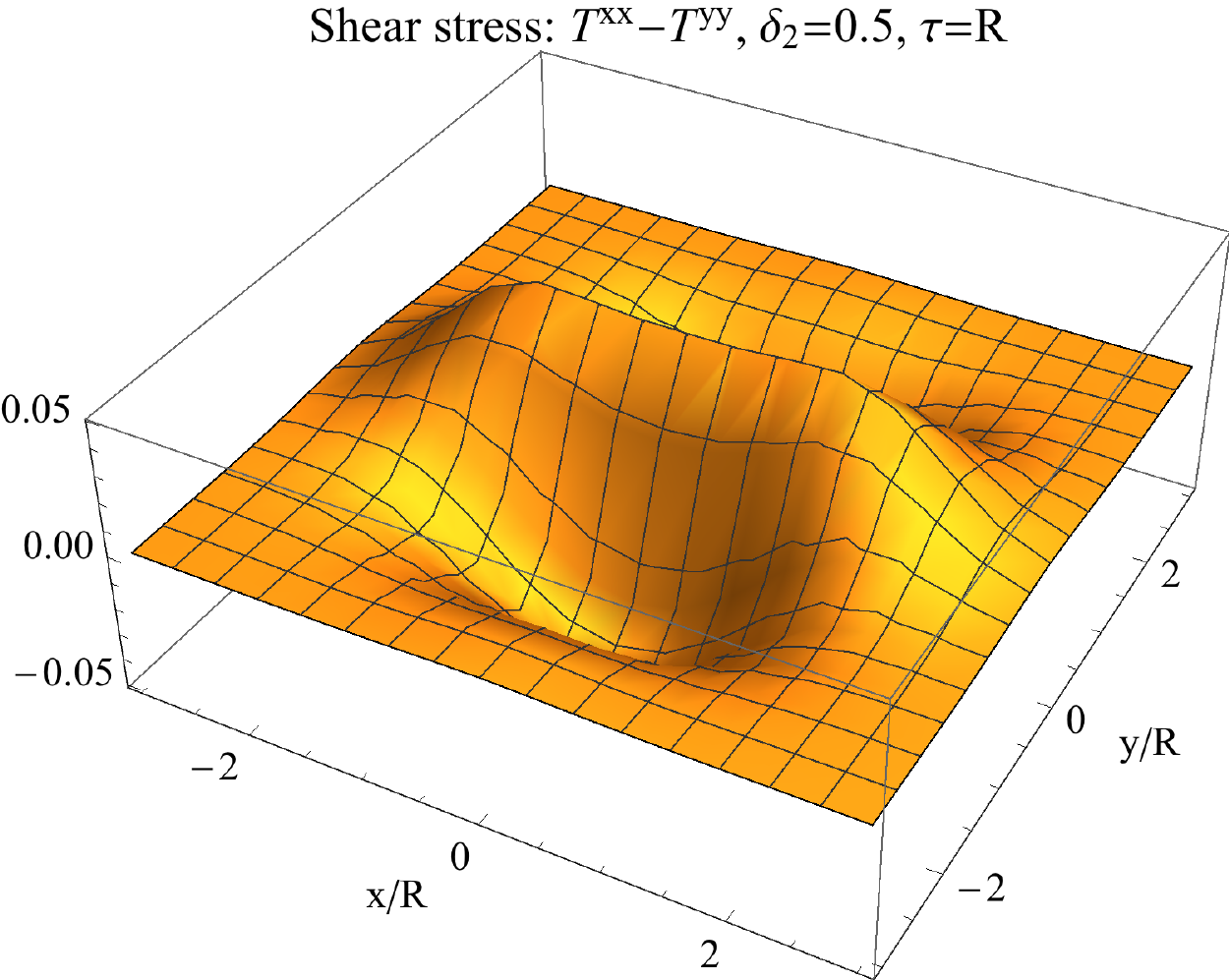}
 \caption{Combining transverse spatial eccentricity, free-streaming, and isotropizing scattering leads to global momentum space anisotropies: (Left) The system starts at $\tau_0$ with an elliptic spatial eccentricity $\delta_2$. The particle production is local and hence momentum distributions at each point are isotropic. (Middle) Free-streaming particles move along the directions of their momentum vectors leading to {\it local} momentum anisotropies. In the central region where most collisions take place, there is an excess of particles moving horizontally compared to vertically moving ones. The interactions in the center region tend to isotropize the distribution function, and thus they reduce the number of horizontal movers and they add vertical movers. This generates a {\it global} momentum anisotropy. (Right) The local momentum space anisotropy measured by $T^{xx}-T^{yy}$ that results from free streaming an initial condition with eccentricity $\delta_2=0.5$
 up to time $\tau = R$.}
 \label{fig1}
\end{figure}

\begin{figure}
\begin{center}
  \includegraphics[width=0.35\textwidth]{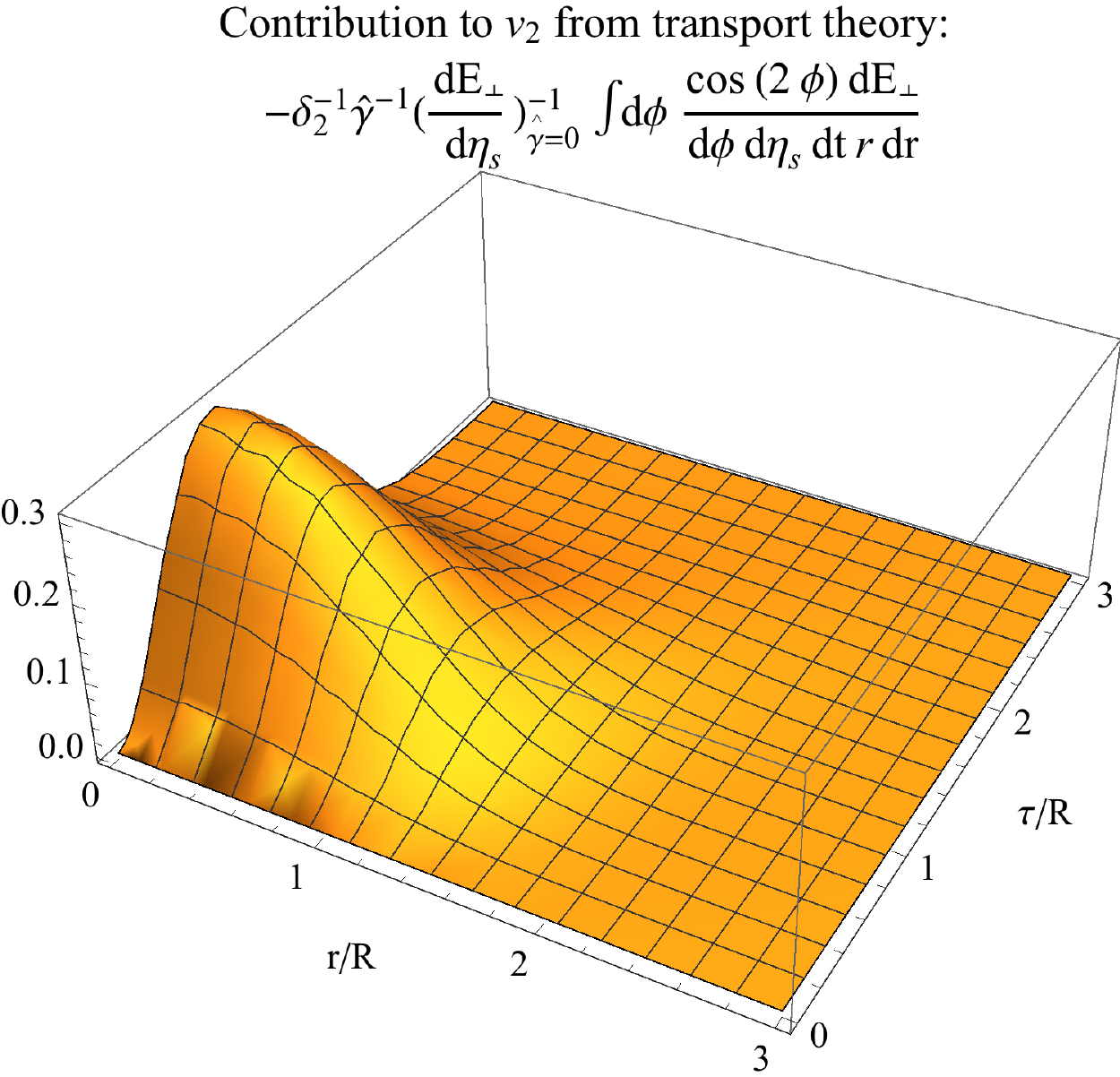}
  \hspace{3cm}
   \includegraphics[width=0.35\textwidth]{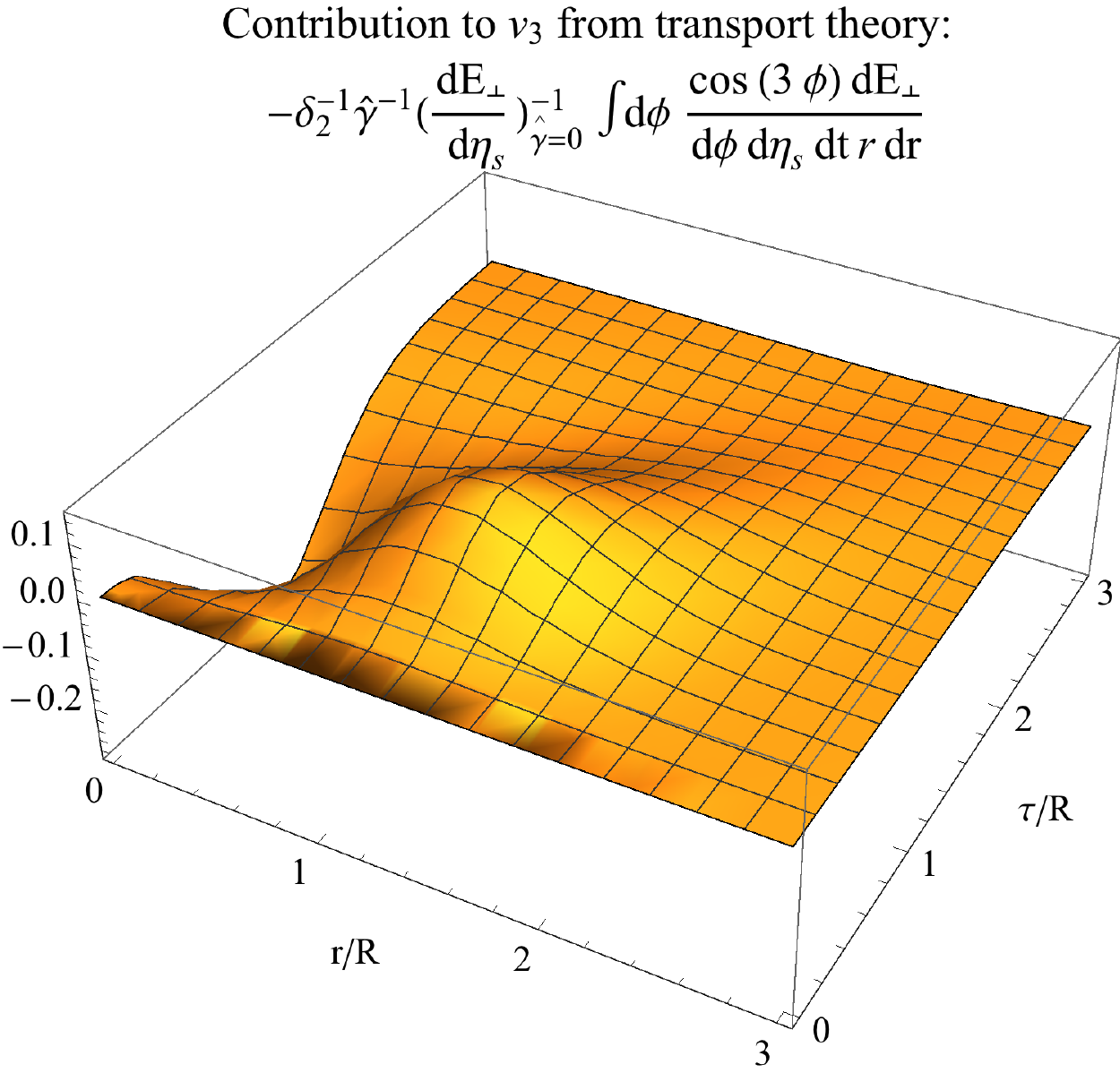}
\end{center}
 \caption{(Left) Contribution to $v_2$ arising from the $\mathcal{O}(\delta_2)$ anisotropy as a function of time. The most of the signal is generated in the center region of the collision where the energy density is the highest at time $\tau \sim R$ when the momentum space anisotropy has developed but before the system has diluted due to radial expansion.  (Right) Contribution to $v_3$ arising from the $\mathcal{O}(\delta_3)$ anisotropy as a function of time.}
 \label{fig2}
\end{figure}

{\bf Propagating eccentricities}.
We want to investigate how azimuthal anisotropies $v_m$ in the transverse flow of energy density and non-linear 
 correlations between these $v_m$'s arise in transport theory (\ref{eq2}) for large mean free path $l_{\rm mfp}$ (i.e. small $\gamma$).
 Our starting point will be an azimuthally isotropic momentum distribution with vanishing longitudinal component $\propto \delta\left(v_z\right)$
 distributed in a spatial profile that is longitudinally boost invariant and composed of an azimuthally isotropic background with
 perturbatively small deviations from azimuthal symmetry 
 \begin{equation}
F(\vec x_\perp, \phi, v_z, \tau=\tau_0) =2 \delta(v_z)\, \frac{(\varepsilon_0 \tau_0)}{\tau_0}  \left[ F_0 (\vec x_\perp,\phi, \tau=\tau_0) + \delta_2 F_{\delta_2} (\vec x_\perp,\phi, \tau=\tau_0) + \delta_3 F_{\delta_3} (\vec x_\perp, \phi, \tau=\tau_0) + \dots \right]\, .
\label{eq6}
\end{equation}
The azimuthally symmetric initial background profile is of Gaussian form,  $F_0(\vec x_\perp, \phi, \tau=\tau_0) = \exp\left[ -(x^2 + y^2)/R^2 \right] $ in the limit $\tau_0 \to 0$. The perturbations $F_{\delta_m}$  are {\it pure} in the $m$-th harmonic, $F_{\delta_2} (\vec x_\perp, \phi, \tau=\tau_0) = (r^2/R^2) \cos(2\theta) F_0(\vec x_\perp, \phi, \tau=\tau_0) = \left(x^2-y^2 \right)
F_0(\vec x_\perp, \phi, \tau=\tau_0)/R^2$ in the limit $\tau_0 \to 0$
and  similarly for the perturbation in $m=3$ with $ r^3 \cos(3\theta) = \left(x^3- 3 x y^2 \right)$. 
Here, $\theta$ ($\phi$) denote the azimuthal angles in coordinate (momentum) space.
Since the integration measure over the normalized longitudinal momentum $-1<v_z<1$ is 
$dv_z/2$, the factor $2$ in front of (\ref{eq6}) implies that $\varepsilon_0 = \varepsilon(\vec{x}_\perp=0,\tau=\tau_0)$ is the initial energy density at the origin.

We want to determine how transport theory in the limit of large in-medium pathlength translates the small spatial perturbations $\delta_m$ into momentum anisotropies. To 
first order in $\gamma \propto 1/l_{\rm mfp}$ (other analyses of this limit are found in Refs.~\cite{Heiselberg:1998es,Kolb:2000fha, Borghini:2010hy, Romatschke:2018wgi}), this amounts to a one-hit dynamics in which the initial distribution 
(\ref{eq6}) is free-streamed up to proper time $\tau$, 
\begin{equation}
F_{\rm free\, stream} (x,y,\phi, \tau) = \frac{\tau_0}{\tau} F(x-(\tau - \tau_0)\cos\phi,y- (\tau - \tau_0)\sin\phi, \phi, \tau_0)\, ,
\label{eq7}
\end{equation}
before entering once the collision kernel. The latter depends on $F_{\rm iso}$, and thus it depends on the velocity $u_\mu$ of the local rest frame and on the comoving energy density 
$\varepsilon$. Both need to be determined from the free-streaming distribution (\ref{eq7}). To this end, we make a tensor decomposition of 
\begin{equation}
T^{\mu \nu} =\int \frac{d^3 p}{(2\pi)^3} \frac{p^\mu p^{\nu}}{p} f = \int_{-1}^{1}\frac{d v_z}{2} \int \frac{d\phi}{2\pi} v^{\mu} v^{\nu} F\, .
\label{eq8}
\end{equation}
For the azimuthally symmetric background distribution $F_0$, the eigensystem of the resulting energy momentum tensor $T_0^{\mu \nu}$ can be given analytically in terms of the harmonic moments $A_m = \int \frac{d\phi}{4\pi} \cos(m \phi)$ $\exp(-(\tau^2+r^2-2\tau r \cos\phi)/R^2)$. In particular,
$\varepsilon_{0} = - \left(A_0 + A_2\right)/2 + A_1/u$ with $u = 4A_1/\left(3 A_0 + A_2 + \sqrt{9 A_0^2 - 16 A_1^2 + 6 A_0 A_2+A_2^2} \right)$, and the background flow field is purely radial $u_\mu= \left(u_0=\sqrt{1-u_r^2}, \right.$ $\left.u_r= u/\sqrt{1-u^2},u_\phi=0,u_z=0 \right)$. Although free-streaming of $F_0$ preserves the 
global coordinate space azimuthal symmetry of the system, the presence of non-zero harmonics $A_m$ in these expressions illustrates that local \emph{momentum space anisotropies} develop over time. This is
so, since  the  free-streaming distribution at any given position $\vec{x}_\perp$ will depend on the direction $\phi$ in which particles propagate. Different positions have different anisotropies that the interactions tend to isotropize with the potential to create imbalance of the particle flow in the final state. In the azimuthally symmetric case the net effect is, however, zero and no anisotropy is generated. In contrast, in the case of an azimuthally anisotropic transverse profile momentum imbalance is generically formed as sketched in Fig.~\ref{fig1}. 

Small spatial eccentricities $\delta_m$ in the initial distribution (\ref{eq6}) lead to computable perturbations of $T^{\mu\nu}$ away from $T_0^{\mu \nu}$. It is then straightforward to
determine the eigensystem of the full energy momentum tensor perturbatively from that of $T_0^{\mu \nu}$ to the needed order in the $\delta_m$'s. For instance, the local energy density of the perturbed
free-streaming solution is of the general form
\begin{equation}
	\varepsilon = \varepsilon_0 + \sum_m \delta_m\, \tilde{\varepsilon}_m \cos\left(m \theta_m \right)
		+ \sum_{m_1,m_2=1}^\infty \delta_{m_1}\, \delta_{m_2} \left( \sum_{\pm} \tilde{\varepsilon}_{m_1m_2,m_1\pm m_2} \cos\left(m_1 \theta_{m_1} \pm  m_2 \theta_{m_2}  \right)  \right)
		+ {\cal O}\left( \delta_{m_i}^3 \right)\, ,
		\label{eq9}
\end{equation}
where the tilde indicates that the angular dependence has been factored out. Here, $\theta_m \equiv \theta - \psi_m$ and equation (\ref{eq9})  is written for a generalization of the initial 
 conditions (\ref{eq6}) for which different harmonic perturbations have different azimuthal reaction plane orientations $\psi_m$, that means 
 $F_{\delta_m} (\vec x_\perp, \phi, \tau=\tau_0) = (r^m/R^m) \cos(m(\theta -\psi_m) ) e^{-r^2/R^2}$. These coefficient functions $\tilde{\varepsilon}_*$ can be determined trivially in explicit form although they are generally lengthy.
For instance 
$$\tilde \varepsilon_2\! =\! \frac{A_1\! \left(4 r^2 u\! +\! r\tau \left(3 u^2\!+\!4\right)\!+\!2 \tau^2 u\right)}{2 R^2 \left(u^2-1\right)} - \frac{A_2\! \left(r^2 u^2\!+\!4 r \tau u\!+\!\tau^2\! \left(u^2\!+\!2\right)\right)}{2 R^2 \left(u^2-1\right)}
+\frac{A_3 \tau u (r u\!+\!2 \tau)}{2 R^2 \left(u^2-1\right)} + \frac{A_4 \tau^2 u^2/R^2}{4-4 u^2} - \frac{A_0\! \left(2 r^2 \left(u^2\!+\!2\right)\!+\!8 r \tau u\!+\!\tau^2 u^2\right)}{4 R^2 \left(u^2-1\right)}\!\, .$$
 We have calculated analogously explicit but lengthy harmonic decompositions of the other eigenvalues and eigenvectors of the full perturbed energy momentum tensor $T^{\mu \nu}
 = T_0^{\mu \nu} + \delta T^{\mu \nu}$ that is obtained from free-streaming of (\ref{eq6}). Armed with this information, we can determine the
 
 {\bf Collision kernel and energy flow.} Free-streaming leads to local azimuthal anisotropies of the spatial distribution, but the experimentally accessible momentum distribution remains 
 azimuthally isotropic. To change the latter, collisions are needed to which we turn now.  We are particularly interested in the measurable angular dependence of the radial flow of energy density
 at late time $\tau_\infty$ when space-time rapidity $\eta_s$ can be identified with momentum rapidity
 \begin{equation}
2\pi \frac{d E_\perp}{d \eta_s d \phi} = \tau_\infty \int 2\pi d^2{x}_\perp  \frac{d T^{0r}}{d\phi}
= \tau_\infty  \int d^2x_\perp \int \frac{2\pi p_\perp d p_\perp}{(2\pi)^2}\int \frac{d p_z}{(2\pi)} p_\perp f(\vec{x}_\perp, p_\perp,p_z,\phi,\tau_\infty)\, .
\label{eq10}
\end{equation}
Since $T^{\mu\nu}$ measures energy-momentum components per unit volume $d^2{x}_\perp\, dz$, the explicit $\tau_\infty$-dependence arises here from 
$dz = \tau_\infty\, d\eta_s$. 
After the last scattering at time $\tau'$, the particles will free-stream from $\tau'$ to $\tau_\infty$.
Since the $\vec{x}_\perp$-integration renders the free propagation trivial, 
$\int dp_z \int d^2 x_\perp  f(\vec{x}_\perp,p_\perp,p_z ,\phi ,\tau_\infty) = \frac{\tau'}{\tau_\infty} \int dp_z \int d^2 x_\perp'  f({\vec{x}'}_\perp,p_\perp,p_z , \phi,\tau')$, we can undo this free-streaming 
and write eq.~(\ref{eq10}) as
\begin{equation}
2\pi \frac{d E_\perp}{d \eta_s d \phi} = \tau_\infty \int d^2 x_\perp \int \frac{dv_z}{2} \sqrt{1-v_z^2}F(\vec{x}_\perp, \phi, v_z,\tau_\infty) = \tau' \int d^2 x_\perp \int \frac{dv_z}{2} \sqrt{1-v_z^2} F(\vec{x}_\perp, \phi, v_z,\tau').
\label{eq11}
\end{equation}
Assuming only a single collision, the correction to the integrated distribution due to the interactions that take place in the infinitesimal time interval from $\tau'$ to $\tau'+d\tau'$ is given by $-C[F_{\rm free\, stream}]d\tau'$. Then integrating over all possible times when the single interaction takes place, we get for the correction of order $\gamma$
\begin{align}
2\pi \frac{d E_\perp}{d \eta_s d \phi}=\frac{d E_\perp}{d \eta_s}\Bigg |_{\gamma=0} -\int d\tau' \tau' \int d^2 x_\perp \int \frac{dv_z}{2} \sqrt{1-v_z^2}C[F_{\rm free\, stream}](\vec{x}_\perp, \phi, v_z,\tau')\, .
\label{eq12}
\end{align}
The collision kernel that enters this one-hit dynamics is known explicitly from (\ref{eq2}) and (\ref{eq4}) in terms of the free-streamed solution $F_{\rm free\, stream}$ obtained from the initial
condition (\ref{eq6}), the energy density $\varepsilon$ calculated as described in (\ref{eq9}) and the rest-frame velocity $u_\mu$ that we have also calculated perturbatively in powers of the 
eccentricities $\delta_m$.
Given that the $\delta_m$-dependencies of $C[F_{\rm free\, stream}]$ are known explicitly, we can Taylor expand in the eccentricities and perform the integrals in (\ref{eq11}), thus obtaining the main result of 
this work
\begin{eqnarray}
 \frac{d E_\perp}{d \eta d \phi}
= \frac{1}{2\pi}\frac{d E_\perp}{d \eta}\Big\vert_{\hat\gamma=0,\delta_n =0} 
&&  \Big\{
1 -0.210 \,\hat\gamma -0.212 \, \hat\gamma \delta_2 \, 2 \cos(2 \phi - 2 \psi_2) 
 -0.140 \,\hat\gamma \delta_3 \, 2 \cos(3\phi- 3\psi_3) \nonumber \\
&&+ 0.063 \,\hat\gamma \delta_2^2 2 \cos(4 \phi - 4 \psi_2) +0.015\, \hat\gamma \delta_2^2 +0.112\, \hat\gamma \delta_3^2 2 \cos(6 \phi - 6 \psi_3) + 0.043 \,\hat\gamma \delta_3^2 \nonumber \\
&&+ 0.088 \,\hat\gamma \delta_2 \delta_3 2 \cos(5\phi - 3\psi_3- 2\psi_2)
 \Big\}\, ,
 \label{eq13}
\end{eqnarray}
where we have followed the angular dependence analytically and performed the remaining $r$- and $\tau$-integrals numerically.

Dimensional analysis of the kinetic theory (\ref{eq2}) reveals that irrespective of the initial conditions, physical results like (\ref{eq13}) depend only on one particular combination 
of system size $R$, initial energy density $\varepsilon_0$  and scale $\gamma$ of the inverse mean free path. 
\begin{equation}
  \hat \gamma =R^{3/4}\gamma (\, \varepsilon_0 \tau_0)^{1/4} \approx R \gamma \left(\, e\, \varepsilon(\vec{x}_\perp = 0,\tau=R)\right)^{1/4} \approx 1.28\,R \,\gamma \, \varepsilon^{1/4}(\vec{x}_\perp = 0,\tau=R)\, .
  \label{eq14}
\end{equation}
Here, the latter form shows 
that the expansion parameter is proportional to the mean free path at time $\tau = R$ when flow is mainly generated; it is obtained
with the help of $\varepsilon(\vec{x}_\perp=0,\tau) = \tau_0\varepsilon_0 e^{-\tau^2/R^2}/\tau + {\cal O}(\gamma, \delta)$. The one-hit dynamics studied here is a truncation of transport theory to first order in $\hat\gamma\sim R/l_{\rm mfp}$. This expansion is justified for $\hat\gamma < O(1)$ but it
breaks down for $\hat\gamma > O(1)$. This is clearly seen from the correction to free-streaming for an azimuthally symmetric distribution (the term $\propto -0.210\hat\gamma$) 
that renders (\ref{eq13}) unphysical for $\hat\gamma > 1/0.210$. This term arises from scatterings that transfer transverse into longitudinal momentum and it indicates that
multiple scatterings need to be resummed for $\hat\gamma$ of this order. 

Within its range of validity ($\hat\gamma < O(1)$), the main qualitative conclusion of (\ref{eq13}) is that both, linear response to spatial eccentricities and non-linear mode-mode coupling are natural 
consequences of a perturbative one-hit dynamics and thus must not be taken for a tell-tale sign of a hydrodynamic mechanism at work. In fact, we could have easily expanded 
(\ref{eq13}) to higher orders in eccentricities to obtain higher non-linear mode-mode couplings (such as $v_6 \propto  \delta_2^3$) or we could have included higher harmonics in the initial condition (\ref{eq6}) 
to find e.g. linear response coefficients $v_4/\delta_4$ and $v_5/\delta_5$. All the linear and non-linear structures observed in the azimuthal distributions of single inclusive hadron spectra 
can be obtained from transport models in the limit of long mean free path. 

A quantitative comparison of the coefficients in (\ref{eq13}) to other model calculations and data is complicated by several issues. For instance, $v_m$ coefficients are typically defined as azimuthal
anisotropies of particle distributions $\frac{d N}{p_\perp dp_\perp\, d \eta d \phi}$. Instead, eq.~(\ref{eq13}) provides anisotropies 
$\frac{d E_\perp}{d \eta d \phi} \propto \frac{d E_\perp}{d \eta} \left[1 + 2 \sum_{m=1} v_m \cos\left(m\phi -m\psi_m\right) \right]$
of the $p_T$-integrated transverse energy flow. Due to the different $p_\perp$-weighting, 
$dE_\perp = p_\perp dN$, the anisotropies thus obtained can be ${\cal O}(30\%)$ larger. Also, the prefactors in (\ref{eq13}) are model-dependent in the sense that they depend 
on the shape of the transverse profile of the initial condition. Only to the extent to which the dynamical response to main spatial characteristics such as $\delta_2$ and $\delta_3$ is robust 
against finer details of transverse profile, can the response coefficients calculated here be taken as indicative of the typical signal sizes obtained from transport theory in the one-hit limit. 
We note in this context that constructions of linear response coefficients by dividing data on $v_m$ with values of $\delta_m$ obtained from model calculations~\cite{ALICE:2011ab,Liu:2018hjh}
show variations of $O(30\%)$ depending on the model from which the eccentricities are calculated. While these considerations caution us that any numerical comparison at face value can
only be indicative of order of magnitude effects, we nevertheless share in the following numerical observations to address the fundamental question whether 
a transport mechanism with significant mean free path can be sufficiently efficient to account for the observed signal size within the range of its validity. 

We base the following numerical comparison on one particularly compact and recent compilation of linear and non-linear response coefficients, see Fig.~1 of Ref.~\cite{Liu:2018hjh}.
This study shows values for $v_2/\delta_2$ ($v_3/\delta_3$) that drop from $\sim 0.3$ ($\sim 0.2$) in the most central (0-5\%) PbPb collisions at 2.76 TeV to values 
$\sim 0.1$ in the most peripheral 70-80\% (in peripheral 40-50\% centrality) collisions. At face value, if we assume $\hat\gamma \sim 1$ in peripheral PbPb collisions of $\sim 50\%$ centrality, then
these findings are consistent with the numbers extracted from  (\ref{eq13}) to leading order in $\hat\gamma$
\begin{equation}
\frac{v_2}{\delta_2} = 0.212 \hat\gamma  \, ,\qquad  \frac{v_3}{\delta_3} = 0.140 \hat\gamma \, .
\label{eq15}
\end{equation}
We note that the factor $(1-0.210\hat\gamma)$ of the zeroth harmonic in (\ref{eq13}) would enter (\ref{eq15}) only to higher order in $\hat\gamma$, where our calculation is not complete. In analogy to (\ref{eq15}), 
one can construct linear response coefficients for central pPb collisions at 5.02 TeV by dividing the measured $v_m$ asymmetries ($v_2 \sim 0.06$, $v_3 \sim 0.02$~\cite{Sirunyan:2017igb})
by estimates of average eccentricities in hadron-nucleus collisions  ($0.25 < \delta_2 < 0.34$, $0.18 < \delta_3<0.32$, ~\cite{Bozek:2013uha}). Again, this lies in the ballpark of (\ref{eq15}) for
$\hat\gamma \sim 1$. Within the above-mentioned uncertainties, this indicates that the range of validity of one-hit dynamics extends to systems as large as those created in 
peripheral PbPb ($\sim 50\%$ centrality) or central pPb collisions. 

For a choice $\hat\gamma \sim 1$,  also the size of the major non-linear mode-mode couplings extracted from (\ref{eq13}) 
\begin{equation}
	\frac{v_4}{\delta_2^2} = 0.063\, \hat\gamma\, ,\qquad \frac{v_5}{\delta_2\, \delta_3} = 0.088 \, \hat\gamma\,  ,
	\label{eq16}
\end{equation}
seem to be, within the stated uncertainties, consistent with the strength of the couplings in peripheral ($\sim 50\%$ centrality) PbPb collision shown
in Ref.~\cite{Liu:2018hjh}, thus reinforcing the qualitative conclusion drawn from the comparison to linear response coefficients. However, the contribution to $v_6$ in (\ref{eq13})
seems to be significantly larger than the value shown in Ref.~\cite{Liu:2018hjh}. We note in this context that while the coefficients in (\ref{eq15}) and (\ref{eq16}) are generated
within a time $\tau \sim R$ comparable to the system size (see Fig.~\ref{fig2}), significant contributions to $v_6$ come in our calculation from times as late as $\tau \sim 4\, R$. 
We refrain from speculating whether these or other reasons are at the origin of the observed discrepancy for $v_6/ \delta_3\, \delta_3$. 

Further qualitative and quantitative features of measured momentum anisotropies  may also be consistent with one-hit dynamics. 
For instance, it has been argued in~\cite{Borghini:2010hy} that the experimentally observed mass ordering of $v_n$'s can also result 
as a generic feature of one-hit dynamics. This is so, because the dynamics leading to $v_n$'s depends essentially on velocities, and this corresponds to a mass ordering in the transverse momenta. 

Since the mean free path $l_{\rm mfp}$ at the typical time $\tau=R$ at which scattering occurs in our calculation is 
$l_{\rm mfp} = \gamma^{-1} \, \varepsilon^{-1/4}(\vec{x}_\perp = 0,\tau=R)$, it follows from (\ref{eq14}) that $\hat\gamma \sim 1$ corresponds to 
$N_{\rm coll} \sim 1$ collision per particle per system size. 
Since $l_{\rm mfp} = \gamma^{-1} \, \varepsilon^{-1/4}(\vec{x}_\perp = 0,\tau=R)$ 
remains almost unchanged with system size while $\hat\gamma$ increases almost linearly with system size, the same transport theory that accounts with $\hat\gamma \sim 1$ for
peripheral PbPb collisions would automatically apply to more
central nucleus-nucleus collisions. It is not the transport theory, but only its one-hit approximation studied here that becomes more questionable for increasing  centrality; 
the intrinsic matter properties of the system remain (almost) unchanged with centrality.

It may be noteworthy that also the AMPT  (see Figs.~1 and 2 of Ref.~\cite{He:2015hfa}) and BAMPS~\cite{Greif:2017bnr} transport codes
simulate central hadron-nucleus collsions with $N_{\rm coll} \sim 1$ collision. Of course, the number $N_{\rm coll}$ or the mean free path 
is not free of model-dependent assumptions about the nature of the collision. In the transport theory (\ref{eq1}), the mean free path $l_{\rm mfp}$ corresponds to the time scale over which a distribution $F$ evolves to the isotropic one $F_{\rm iso}$.
If instead a more sophisticated collision kernel dominated by small-angle scatterings is implemented in a transport code, a larger number of collisions may be needed for isotropization. Thus, in any given transport code, the efficiency to isotropize will depend on the microscopic dynamics (whose specification introduces some model-dependence); and in the present transport theory (\ref{eq1}), this isotropization efficiency is parametrized in terms of one single parameter $\gamma$, irrespective of how it arises microscopically. 
There is no reason that eq.~(\ref{eq1}) must arrive at the same number of collisions than a transport code
(that may supplement transport mechanisms by other model-dependent features such as string melting, hadronization prescriptions, etc),
but the study of (\ref{eq1}) teaches us that there is a class of transport models that can generate within the range of their validity a distribution (\ref{eq13}) by a microscopic dynamics that realizes a particular value of $\gamma$. 

For transport theory to apply,  the mean free path of quasi-particles must be larger than the typical size of their 
wave-packets~\cite{Danielewicz:1984ww},
$l_{\rm mfp} > 1/T$. Here, we have argued that $\hat\gamma \sim 1$ describes peripheral (50\% centrality) PbPb collisions
which corresponds to a mean free path of the order of the transverse size of these collisions and which is significantly larger
than the expected inter-particle distance $\sim 1/T$ in PbPb collisions. This supports the applicability of transport theory to 
ultra-relativistic pA and AA collisions. Indeed, the relation $l_{\rm mfp} = 5 \frac{\eta}{s T} > 1/T$ suggests that
transport theory may be applicable to systems of small shear viscosity over entropy ratio, as long as $\eta/s \gtrsim 0.2$.
To extract the shear viscosity over entropy density ratio $\eta/s$ for 50\% central PbPb collisions described by $\hat\gamma \sim 1$,  
we use that the transverse energy produced in such collisions is set by measured quantitities,
$\frac{dE_\perp}{d \eta}=\langle p_\perp \rangle \langle \frac{d N}{d \eta}\rangle \approx 100$ GeV \cite{Aamodt:2010cz}  and their system size $R$ can be estimated from the nuclear
overlap function at impact parameter $\sim 10$ fm \cite{Kolb:2001qz}, which is $R \sim 3$ fm.
In our set-up, $\frac{dE_\perp}{d \eta} = \pi R^3 e \varepsilon(R) $ converts these values to 
an energy density at time $\tau = R$ that takes the value $\varepsilon(R) \approx (240 {\rm MeV})^4$. This leads to $\frac{1}{\gamma} = e^{1/4} R \varepsilon^{1/4}(R) \approx 4.6$ and, given that the above
estimates were taken on the conservative side,  this translates according to eq.~(\ref{eq5}) into a shear viscosity 
\begin{equation}
\eta/s = \frac{0.11}{\hat \gamma} \left( \frac{R}{\pi} \frac{d E_\perp}{d\eta}\right)^{1/4}\approx  0.5
\label{eq17}
\end{equation} 
or larger. 
Uncertainties about the centrality class for which $\hat\gamma\sim 1$ enter this expression only via the fourth root
of the system size and the transverse energy in that centrality class. This makes the above estimate robust. 
We recall that for a kinetic transport description,
$\eta/s$ is a derived quantity that requires specifying the relation between energy density and temperture which does not
enter the dynamical evolution. In the BAMPS transport model with parameter settings that reproduce data from nucleus-nucleus 
collisions at RHIC and at the LHC, a corresponding (somewhat temperature-dependent) value for $\eta/s$ was obtained~\cite{Uphoff:2014cba} that
in the low temperature range in which flow is built up is indeed roughly consistent with the estimate (\ref{eq17}). 
We are unaware of other transport model comparisons to data that quote values of $\eta/s$. 
We note that the estimate (\ref{eq17}) is a factor 6 larger than the perfect fluid limit $\eta/s = 1/4\pi$. Also, 
remarkably, evaluating the perturbative QCD result~\cite{Arnold:2000dr, Arnold:2003zc,Ghiglieri:2018dib} with realistic values 
for the QCD coupling constant results in a value for $\eta/s$ of similar magnitude. While the value of $\eta/s$ supported 
by the present study is still exceptionally small compared to that of common substances like helium, nitrogen or water~\cite{Kovtun:2004de}, 
it suggests that the systems created in nucleus-nucleus and proton-nucleus collisions may differ from the perfect fluid paradigm 
in that they allow for quasi-particles of significant mean free path. 

We acknowledge helpful discussions with Ulrich Heinz.

\bibliographystyle{elsarticle-num}

\end{document}